%% file: main.tex
\documentclass[a4paper,UKenglish,draft]{dagman-v2021}

\usepackage{microtype}
\newcommand{\centertitle}[1]{\noindent{\begin{center} \bf \large #1\end{center}}}
\usepackage{fancyhdr}
\fancypagestyle{firstpage}{%
}

\newcommand{\ie}{\textit{i.e.}}
\newcommand{\eg}{\textit{e.g.}}
\newcommand{\principle}[1]{\textsf{#1}}

\subject{Manifesto from Dagstuhl Perspectives Workshop 24452}
\title{Reframing Technical Debt}
\titlerunning{Reframing Technical Debt}

\editor{Paris Avgeriou}{University of Groningen, Netherlands}{p.avgeriou@rug.nl}{}{}
\editor{Ipek Ozkaya}{Carnegie Mellon Software Engineering Institute, USA}{ozkaya@sei.cmu.edu}{}{}
\editor{Heiko Koziolek}{ABB Corporate Research, Germany}{heiko.koziolek@de.abb.com}{}{}
\editor{Zadia Codabux}{University of Saskatchewan, Canada}{zadiacodabux@ieee.org}{}{}
\author{Neil Ernst}{University of Victoria, Canada}
{nernst@uvic.ca}{}{}

\authorrunning{Avgeriou \emph{et al.}}


\seminarnumber{24452}
\semdata{03.--08.~November, 2024 -- \url{https://www.dagstuhl.de/24452}}


\seminarnumber{24452}
\semdata{November 04--08, 2024 -- \url{https://www.dagstuhl.de/24452}}

\volumeinfo
  {John Q. Open and Joan R. Access}
  {2}
  {A Manifesto Sample}
  {1}
  {1}
  {1}
\DOI{10.4230/DagMan.1.1.1}

\begin{document}

\maketitle
\additionaleditorsHeading{Apostolos, Ampatzoglou 
 Lodewijk Bergmans, 
 Markus Borg, 
 Alexandros Chatzigeorgiou, 
 Marcus Ciolkowski, 
 Stefano Dalla Palma, 
 Florian Deißenböck, 
 Philippe-Emmanuel Douziech, 
 Daniel Feitosa, 
 Michael Felderer, 
 Collin Green, 
 Ciera Jaspan, 
 Ron Koontz, 
 Christof Momm, 
 Brigid O'Hearn, 
 Klaus Schmid, 
 Carolyn Seaman, 
 Tushar Sharma, 
 Guilherme Horta Travassos,
 Roberto Verdecchia, 
 Marion Wiese}
 \thispagestyle{firstpage}
 \include{bvpone}
 \newpage

\tableofcontents

\section{Introduction}
Technical Debt is widely recognized as a critical software engineering concern, imposing significant costs on productivity, maintainability, and evolvability~\cite{Kruchten2012}. Developers reportedly spend a substantial portion of their effort on Technical Debt-related activities~\cite{Besker2018}. Current approaches to combat Technical Debt often focus on low-level code analysis and patchwork solutions, while broader socio-technical and architectural considerations receive less attention~\cite{Ernst2015,Avgeriou2021,Jaspan2023,Biazotto2024}. We also continuously strive to evolve how we collect, interpret, and utilize data to effectively manage Technical Debt and create future value. 

In this Perspective Workshop, we did not attempt to redefine the term `Technical Debt.' Instead, we adopt the Technical Debt definition articulated at the Dagstuhl Seminar 16162 on this topic which, through consensus, has stood the test of time \cite{DagstuhlAvgeriou2016}:

\emph{
In software-intensive systems, Technical Debt is a collection of design or implementation constructs that are expedient in the short term but set up a technical context that can make future changes more costly or impossible. Technical Debt presents an actual or contingent liability whose impact is limited to internal system qualities, primarily maintainability and evolvability.}

Furthermore, we adopt the definitions of two more terms from the Dagstuhl seminar 16162~\cite{DagstuhlAvgeriou2016}: the \textbf{principal} of Technical Debt refers to ``the cost that it would now take to develop a different or better solution'', thus removing the aforementioned liability; the \textbf{interest} of Technical Debt refers to ``additional cost incurred by the project in the presence of technical debt, due to reduced velocity (or productivity), induced defects, and loss of quality.'' The interest continuously accumulates until one of two outcomes occur: 1) the relevant technical debt is resolved or 2) the environment changes, resulting in the original debt no longer being relevant (e.g., a code unit is no longer touched). 

All software systems contain Technical Debt where Technical Debt can be introduced across the software development lifecycle. Earlier lifecycle Technical Debt typically imposes broader impact on software systems as compared to Technical Debt introduced late in the development cycle. The scope of this manifesto, aligned with the above definition of Technical Debt, includes all relevant artifacts and processes within a \textit{Software Engineering} lifecycle from cradle to grave: from requirements to design and implementation, to testing and continuous integration and continuous deployment (CI/CD), and during maintenance and evolution iterations. In the development of software-intensive systems (e.g. embedded systems, automotive, avionics), Technical Debt also exists at the system and hardware levels, however, only software engineering Technical Debt is within scope of this manifesto.

This document begins by outlining our shared values, beliefs, and principles. These comprise our vision for a new framing of Technical Debt Management thereby establishing it as a routinely followed and well-defined software engineering practice. This document then briefly sketches the current landscape in Technical Debt Management (\ie, where we are now), to understand the distance from the vision (\ie, where we want to be), and concludes with a roadmap to realize the vision (\ie, how to get there).

\section{Values, Beliefs, and Principles}

\subsection{Values}
Values reflect what matters the most.\\

\noindent \textbf{V1 - We value psychological safety and trust between technical and business stakeholders for effective Technical Debt Management.} \label{V.safety-trust}

Psychological safety\footnote{For a canonical reference of the term, please see \cite{doi:10.2307/2666999}.} is essential for effective Technical Debt Management as it implies that there is no risk of retribution. Instead, people can talk openly about Technical Debt and how it was introduced without concern about negative career impact.
Equally important is building trust between technical and business stakeholders: they both have legitimate concerns about the software being developed, and they need to mutually understand and respect these concerns by working together to address them. Accountability translates into appointing concrete roles to manage Technical Debt instead of blaming other stakeholders.
\\

\noindent \textbf{V2 - We value simple, actionable, value-based communication of Technical Debt to all stakeholders over excessive, minute, overwhelming details.}
\label{V.comm}

When communicating Technical Debt information to the development team and even more so to management, clear communications and supporting information should be minimally complex and focus on actionable points. Although metrics, data, and trend plots have a role in effective Technical Debt Management, an overload of fragmented information may confuse rather than enable prioritized decision-making.
Communication, especially with management, should also emphasize information that impacts business value more than any single metric or set of metrics. The key is not to over-simplify Technical Debt information, but instead to foster a practical and actionable Technical Debt Management approach that is clearly understood by developers and management.
\\

\noindent \textbf{V3 - We value transparency, explainability, and replicability in the identification, measurement, and prioritization of Technical Debt.}
\label{V.transparency}

The methods by which Technical Debt is identified, measured, and prioritized should not be obscured behind complicated tooling, esoteric rules, and undisclosed data sets. Instead, Technical Debt items should be traced to specific data sources, where measurement and prioritization rules should be open and explainable to all stakeholders.
\\

\noindent \textbf{V4 - We value both objective and subjective data collection along with research methods to identify, measure, and prioritize Technical Debt over single-method approaches.}
\label{V.data}

Technical Debt items may take many different forms, therefore, identifying and measuring them requires incorporating multiple mono- or inter-disciplinary approaches (e.g., from economics or psychology), metrics, and benchmarks that might differ for each form of Technical Debt depending on the context. Relevant data may be of several different types, ranging from quantitative metrics to qualitative data (e.g., open-ended survey responses and interviews that capture opinions on the severity/priority of Technical Debt items or the probability of specific components changing in the future).
\\

\noindent \textbf{V5 - We value software architecture understanding across the team as part of effective Technical Debt Management}.
\label{V.architecture}

A precondition to understanding Technical Debt items, their consequences, and the effects of repaying them is understanding the current architecture and how it has evolved over time. The entire team (developers, managers and other stakeholders) should share such an understanding to enable them to align and check the conformance of the architecture implementation. This should also occur in parallel with maintaining code quality, and it often results in improving it. Teams should consider architecture understanding as an essential part of development competency as well as a means to reason about business value and technical tradeoffs. 

\subsection{Beliefs}
Beliefs are what we consider to be true, even without formal proof.\\

\textbf{B1 - We believe that sustainable software delivery requires proactive and continuous Technical Debt Management.}

Technical Debt Management is not a one-time activity but requires continuous effort. It is a \textit{culture} that permeates the entire software development lifecycle and system evolution in terms of the application of processes and tools.
Technical Debt Management is effective when it is explicitly and systematically recognized, valued, and rewarded, and is a part of ongoing resource planning activities.
\\

\noindent \textbf{B2 - We believe the most effective Technical Debt Management is as automatic as possible and as manual as needed.}

We value Technical Debt Management automation to the greatest extent practicable, \eg, as part of a general DevOps toolchain, continuously analyzing, measuring, and updating relevant data. Concurrently, we recognize the need for human-in-the-loop approaches, with human insights needed for data interpretation, integration, and decision-making based on the results of automated processes. Therefore, humans must perform manual supervision and intervention informed by (often informal) contextual factors that are essential to formulate effective actions supported by automation.
\\



\noindent \textbf{B3 - Technical Debt must be managed irrespective of how the software system is created, including generation of code and other artifacts by AI-assisted tools}.

Evolving technologies, such as the emergence of AI-generated artifacts, will not eliminate the need for Technical Debt Management, although it may require adjustments to the process. One cannot assume that tool-generated implementation or design artifacts have no Technical Debt. The entire set of software development artifacts should be considered when evaluating and measuring Technical Debt.
\\

\noindent \textbf{B4 - We believe that items that developers observe as Technical Debt should be addressed, even when they are not supported by metrics}. 

Some Technical Debt is objectively identifiable and measurable, while other Technical Debt is subjectively identifiable (but its interest is still measurable). The two sets partially overlap, \ie, human perceptions of Technical Debt are not necessarily correlated to metrics and should be taken into account alongside metrics. Human perceptions should be prioritized and handled, e.g. contextual knowledge, such as inferior or outdated technology choices, is typically not reported in today's Technical Debt metrics but is often well-understood by software developers. The inability to provide quantitative data on Technical Debt principal and interest must not impede steps toward refactoring based on well-known best practices. It can not unnecessarily lead to undue emphasis on polished code without addressing underlying issues or significantly improve development velocity.
\\

\noindent \textbf{B5 - We believe that not all issues identified by stakeholders or tools are Technical Debt}. 

We strongly caution against the over-use or misuse of the term Technical Debt. Not everything detrimental to a software engineering project is Technical Debt. We have observed that broad usage of the term Technical Debt for many kinds of problems in software development projects has led to confusion and has complicated constructive measures towards effective Technical Debt Management. Additionally, tools may identify violations or other issues that are not Technical Debt, for example, if they do not incur interest.

\subsection{Principles}
Principles are guidelines derived from the stated values and beliefs toward actions and decisions.\\

\noindent \textbf {P1 - Share Responsibility for Technical Debt Management.}

We advocate for establishment of shared responsibility and mutual understanding among business and technical stakeholders. Although designated roles may oversee execution, both business and technical stakeholders should participate to define and execute the Technical Debt Management process, where they both share ownership. This is not a one-off activity but a process of continuous learning, evolution, and collaboration of the software stakeholder community in Technical Debt Management. Finally, designating champions for Technical Debt Management within software development teams is critical to fostering consistent and mutual understanding among stakeholders, raising awareness, promoting shared responsibility, and facilitating training. \\

\noindent \textbf {P2 - Manage Technical Debt in Alignment with its Context.}

Technical Debt is relevant only within a given context. Teams need to select metrics, collect data, and interpret them quantitatively and qualitatively with respect to that context. 
In particular, the relative value of eliminating Technical Debt (paying the principal now) versus investing elsewhere (while accepting current and future interest payments) is highly dependent on context (\eg, system size and age, criticality, business model, team distribution, governance, rate of change, impact on velocity, cost to fix). Context makes Technical Debt comparison across projects difficult, if not impossible. \\

\noindent \textbf{P3 - Collect Comprehensive Data for Technical Debt Management.}

An effective Technical Debt Management approach collects and triangulates data from multiple sources, including software engineering artifacts, processes, and people, and both qualitative and quantitative data sources. Teams need to collect data that span all activities, from identifying to monitoring Technical Debt, and go beyond what is convenient to collect. While static code analysis tools may seemingly offer a quick way to reveal Technical Debt, they provide a limited view and must be accompanied by a focus on architecture. Additional developer and management perspectives are required to review relevant contextual factors. \\

\noindent \textbf{P4 - Avoid a One-Size-Fits-All Technical Debt Metric.}

Individual stakeholders require different Technical Debt metrics. 
Teams should opt for a set of carefully thought-out metrics over single, narrowly scoped, non-actionable metrics. The set of metrics should be re-evaluated and revised periodically to maintain alignment with trending data and mitigate concept drift. Reusable metrics should be properly validated against benchmarks, and against developer perceptions, for stability over time, and for generalizability across organizations.\\

\noindent \textbf{P5 - Build Seamless and Integrated Technical Debt Management Toolchains with Human Oversight.}

Software development teams need to select and build Technical Debt Management toolchains integrated within existing workflows, development tools, infrastructures, and processes. Tool vendors need to provide adequate support to achieve this level of integration. While integrated and automated approaches are desirable to avoid overburdening teams and to improve consistency, teams need to ensure a human is in the loop to interpret data, and make decisions (with minimal additional effort), as fully automated approaches at scale and level of correctness desired do not exist. To optimize human intervention, tools should be configurable to accommodate the needs of users with different levels of knowledge and experience.\\

\noindent \textbf{P6 - Make Technical Debt Visible.}

We advocate documenting both deliberate and inadvertent Technical Debt items, including the principal (cost to pay back), value (benefit of the decision), and interest (cost of change) of each item, to enhance transparency to both business and technical stakeholders. Recording the Technical Debt item cause and the respective decision on how to resolve that item enables effective Technical Debt Management. However, Technical Debt documentation must not lead to heavy-weight, bureaucratic practices that unnecessarily overburden developers; instead, it must focus on the most significant and impactful aspects, such as the root cause.\\

\noindent \textbf{P7 - Elevate the Role of Architecture in Technical Debt Management.}

Taking an architectural perspective (including a focus on business value) in Technical Debt Management significantly impacts identifying, resolving, and preventing Technical Debt compared to code or other artifacts.  This impact is often underestimated. Tools and practices must consider architecture to balance the perspective on code. \\ 

\noindent \textbf{P8 - Develop Fit-for-Purpose Technical Debt Benchmarks.}

We consider that purpose-built Technical Debt benchmarks (\ie, standardized tests) are significantly distinct from classical static code analysis benchmarks. Such purpose-built Technical Debt benchmarks should not only focus on what can be easily extracted with current tools but also cover contextual factors as well as architectural concepts that are important for the identification and prioritization of Technical Debt. There is a benefit to defining different benchmarks for Technical Debt based on the purpose, such as governance, tool comparison, or application domains, the inclusion of diverse software artifacts, and a variety of development contexts. This will not only boost appropriate tool development, but it will also assist practitioners in adopting tools. \\  

\noindent \textbf{P9 - Make Intentional Technical Debt Tradeoff Decisions.}

An effective Technical Debt Management approach bases the analysis of Technical Debt tradeoffs on data and transparent decision-making. All stakeholders must recognize that Technical Debt Management involves intentional compromise (\eg, prioritizing among features, defects, security vulnerabilities, and Technical Debt items). Especially at the moment that Technical Debt is incurred, the involved stakeholders should document the decision, the tradeoff and a remediation plan. \\

\section{Current Landscape}
The current state of Technical Debt Management practice indicates significant progress in some areas but reveals some misconceptions and gaps that can potentially be resolved if the stated Principles are applied. In addition, as software engineering research and technologies evolve, new areas of investigation will emerge, which may drive the need to revisit the stated Values, Beliefs, and Principles (see the roadmap in Section \ref{sec:roadmap}). We summarize the current landscape of Technical Debt Management research and practice from the perspectives of value-creation, Technical Debt tooling, data collection, the role of architecture, and socio-technical aspects. 

\subsection{Technical Debt as Value-Creation}

In certain situations, taking on Technical Debt can yield significant benefits, sometimes even outweighing the costs of paying the interest. As outlined by Fowler~\cite{Fowler2009}, developers may take on Technical Debt deliberately and prudently to obtain value, \eg, to meet a business goal. Self-admitted Technical Debt~\cite{sierra2019survey}, referring to developers exposing Technical Debt in code comments as notes to their teammates or themselves, can also manifest this phenomenon. While many methods and tools already address inadvertent and reckless Technical Debt~\cite{seaman2011measuring,alves2016identification,Avgeriou2020}, a balanced Technical Debt Management approach~\cite{li2015systematic} could integrate methods and tools for deliberate and prudent creation of Technical Debt~\cite{borup2021deliberative}, for example assessing the potential value such Technical Debt could add despite its long-term costs to inform decision making.

\subsubsection{State of the Art}

Typical value assessment methods, which are also routinely used in software development projects, include Net Present Value (NPV), return on investment (ROI) or real options analysis (ROA)~\cite{jones2007investments}. However, product managers rarely use these methods in Technical Debt Management~\cite{pichler2010agile,kruchten2019managing}. In addition, software developers are typically not familiar with them. 

The state of the art on the value of Technical Debt includes some proposals~\cite{schmid_formal_2013}, for example estimating the costs that refactorings of specific TD items could save in different future evolution scenarios. However, these are hardly applied in practice today, since they require careful planning and a good understanding of TD items in a system.
Quantification of Technical Debt has received considerable research, focusing on costs and impact to source code~\cite{nayebi2019longitudinal,perera2024systematic}, while often neglecting a value quantification that requires considering soft factors, such as the market situation or the financial impact of specific features~\cite{schmid_formal_2013}. Furthermore, the consequences of a design decision (both in creating value and incurring extra maintenance costs) may only become apparent at a later stage, thereby complicating an accurate analysis. However, a balanced approach to Technical Debt Management would consider both benefits and drawbacks of Technical Debt based on the available information, avoiding short-term reckless behavior as well as unnecessary and inefficient ``gold-plating''~\cite{kruchten2012technical}.

\subsubsection{Challenges}

To reason about the (net) value of a Technical Debt Item (TDI), practitioners require tools and approaches for the following categories of value: 
\begin{enumerate}
    \item Value from taking out the loan: (short-term) benefits of the TDI.
    \item Value from paying the interest: tolerate maintenance costs by postponing the resolution of a TDI in favor of prioritizing backlog items that provide higher business value.
    \item Value from paying back the principal: the cost of fixing the design decision that constitutes the TDI. 
\end{enumerate}

\noindent \textbf{Value from Taking Out the Loan}\\
Developers often accept Technical Debt to avoid creating and implementing a more elaborate or involved design, and thereby deliver sooner. This may be useful to, for example:
\begin{itemize}
    \item Meet a business-critical deadline (\eg, a contractually agreed release date)
    \item Achieve shorter time-to-market (\eg, success may depend on early market introduction)
    \item Obtain earlier feedback on how to improve the product (\eg, refining requirements from user comments)
    \item Learn about technical risks (\eg, 'fail fast' to avoid dead-ends)
    \item Avoid over-investment through over-engineering of an improvement that has narrow or limited-time value (\eg, building a bespoke feature for a critical customer as opposed to building that feature to serve multiple customers)
\end{itemize}

\noindent Typically, taking on technical debt results in reduced time or effort and thus immediate cost savings, for example, by deferring proper design or testing. As most software development teams are cost-constrained and may additionally be time-constrained by external deadlines and fixed scope, in this case, taking on Technical Debt can be the only option to release a product, thus bringing value.\\

\noindent \textbf{Value from Interest Payment}\\
Taking on Technical Debt usually leads to increased cost of change in future development. In cases where these costs are low and where other backlog items suggest higher value, these costs may sometimes be justified and provide indirect value. The extent of these costs depends on several contextual factors. For a comprehensive understanding, the following distinct aspects need consideration:
\begin{itemize}
    \item Maintainability: cost of change over time, \ie, the key consideration of Technical Debt.
    \item Indirect cost of lower delivered quality: operations cost and risks (downtime, security, laws, \dots), defects and user satisfaction.
\end{itemize}

\noindent Note that long-term costs (interest) are heavily dependent on system evolution and its usage, and hence sometimes difficult to predict~\cite{schmid_formal_2013}. For example, the Technical Debt interest in a module that is lightly touched after initial development may be ignorable until conditions change. In other cases, paying a small interest may be favorable over the efforts for paying back the principal, and thus bringing value to a project.\\

\noindent \textbf{Value from Paying Back the Principal}\\
The costs of paying back principal can be considered a refactoring or renovation effort; for example, the effort required to undo a poor design decision. There are several factors that influence these costs:
\begin{itemize}
    \item How widespread is the impact of the Technical Debt item; \eg, a wrong or uninformed architectural decision can have an impact across the system.
    \item Correspondingly, the more time that passes before a TDI is addressed, the larger the impact that TDI can have, as more code is influenced and needs to be revised.
    \item How well tested is the code to be refactored? This can make a large difference in both the effort and the risk of refactoring.
\end{itemize}

\noindent While the importance of these three categories of value is currently understood, practices and tools to better operationalize them are missing. 
Furthermore, while software teams are aware of the tradeoff between these different values, they have no means to track this tradeoff. Finally, software engineers and other decision makers are currently lacking effective ways to link these values with traditional business metrics (\eg, depreciation, asset/liabilities, balance sheets).

\subsection{Tooling for Technical Debt Management}
\label{sec:NextGenTooling}

Tools play an important role in identifying and managing Technical Debt, especially given the size, scope, and complexity of artifacts in modern software development processes. This section summarizes the current landscape of tools related to Technical Debt and its management, highlighting what they offer, how they can be effectively used, their gaps, and what the software engineering community, specifically the industry, expects from them in the future. It is important to clarify that while many excellent tools play a fundamental role in helping organizations manage software development and quality and may include features to assist in Technical Debt Management, currently, the software industry does not have common agreement on what essential tool features are needed for Technical Debt Management. Avgeriou and colleagues studied features from 26 tools~\cite{Avgeriou2020} and concluded that different tools adopt different terms, metrics, and ways to identify and measure technical debt. This results in confusion among software engineers and leads to inconsistencies in effective quantification and management of Technical Debt~\cite{Avgeriou2020}. Specialized research tools for Technical Debt Management exist, \eg, VisminerTD~\cite{Mendes2019} and Anacondebt~\cite{Martini2018a}; however, they they frequently do not scale to the level expected by industry. 

\subsubsection{State of the Art}

Tools that can help analyze Technical Debt span a wide spectrum, ranging from syntactical linters to metrics and static analysis tools, including code smell detection tools~\cite{Sharma2018}. Tools in this space can broadly be categorized as: a) tools that identify code quality issues (including smells and metrics) that indicate the presence of Technical Debt; and b) tools that attempt to determine a broad proxy component of code quality issues by incorporating developer time to resolve those issues into the analysis. Examples of tools in this second category include: SonarQube (identifies code quality issues and assigns a Technical Debt value to them based on predefined rules)~\cite{SonarQube}, CodeScene (assesses code quality based on code smells and presents the aggregated result to the user as CodeHealth\texttrademark)~\cite{CodeScene}, Designite (computes common code quality metrics and detects code smells at implementation, design, and architecture granularities)~\cite{Designite}, CAST (offers a software analytics platform including Technical Debt insights)~\cite{Cast}, and TeamScale (focuses on end-to-end code quality spanning code, tests, requirements, and architectural aspects)~\cite{TeamScale}. \\

\noindent \textbf{Tools Supporting Detecting and Measuring Technical Debt}\\
Detection and measurement usually go together, as some forms of detection are performed through metrics, which can also further assist in prioritizing different Technical Debt items. The state of the art and practice in this space is fragmented, where most tools rely on existing established metrics. Current tools provide traditional code quality metrics to measure complexity (\eg, cyclomatic complexity), cohesion (\eg, lack of cohesion among methods (LCOM)), and coupling (\eg, Fan-in and Fan-out) of source code elements such as methods and classes. Several tools offer code quality issue detection, including code smells at different granularities and different artifacts (such as for test and infrastructure scripts in addition to source code). In recent years, tool providers have expanded their offerings by connecting tools with developer workflows such as CI/CD pipelines. The visualization aids offered by some of these tools help teams consume Technical Debt identification information and follow the trend of their projects code quality. While all of this information is essential in managing the quality and evolution of a software system, by itself, it is insufficient to assess Technical Debt. Most of these tools focus on code artifacts, sometimes including build and test aspects. 

Technical Debt may be identified in one development artifact (\eg, a security issue requiring frequent patches), but resolving it often involves examining other artifacts (for instance, not just analyzing the code but also investigating components through which the security concern could propagate via flawed APIs). While categorizing Technical Debt based on the artifacts current tools can measure (\eg, security debt, requirements debt, code debt) can help teams focus their analysis, it can also lead to tunnel vision---causing them to overlook harder-to-identify Technical Debt like architecture-related issues which often span multiple development artifacts.   \\

\noindent \textbf{Tools Supporting Technical Debt Documentation and Communication}\\
Industry experiences have led researchers and tool vendors to advocate for increased Technical Debt visibility to ensure its management, suggesting approaches such as including Technical Debt items in the backlog~\cite{kruchten2019managing}. Issue tracking platform tools (\eg, Jira~\cite{Jira}) accomplish this and include features to define custom labels which explicitly refer to TDIs. Self-Admitted Technical Debt research has emphasized the importance of linking identified debt in a code base to its management by automatically logging it to issue management platforms~\cite{KASHIWA2022106855}.  However, tuning issue-tracking tools to accommodate Technical Debt Management (e.g., by adding an issue category dedicated to Technical Debt) must be compatible with an organization's use of the issue-tracking tool, as such use varies considerably across organizations.

Visualization features help prioritize and manage Technical Debt. For example, SonarQube includes a heat map differentiating areas in a codebase with many issues, including potential Technical Debt. 
Technical Debt research has emphasized the creation and review of dashboards/visualizations for the management and prioritization~\cite{Kaiser2011, Mendes2019} of Technical Debt. However, in practice, it isn't clear that software engineers accept and use these features. 

Existing code quality tools for analyzing Technical Debt and taking advantage of issue trackers for Technical Debt Management have value. However, tools and features that take a system's context into account and assist in detecting, prioritizing, resolving, and propagating changes to the overall view of a system's Technical Debt are missing. 

\subsubsection{Challenges}

Despite significant progress, existing tools have room for improvement in supporting end-to-end Technical Debt Management with better context alignment. Generic tool support for Technical Debt Management is also challenged by the wide variability in development projects in terms of size, cost and schedule constraints, complexity, ``inherited'' debt from reused software, safety level, level of Technical Debt Management skills, customer expectations, and team overall experience level. To improve tooling for Technical Debt Management, the following challenges must be addressed.\\

    \noindent \textbf{Separate Code Quality from Technical Debt} \\
    Identification of code quality issues is the first step towards Technical Debt awareness for some kinds of Technical Debt. However, not all code quality issues contribute to Technical Debt and code may not even be the most appropriate artifact to assess some forms of Technical Debt form. \textit{Context} is the difference between a code quality issue, such as a code smell, and a TDI. If a code quality issue increases the cost of change only then is it considered as a TDI. Most current tools do not, however, provide automatic, or even semi-automatic, support to proactively identify the context-specific cost of change and priority of change due to a code quality issue. Hence, automated detection of Technical Debt often treats all code quality issues equally. \\
    
    \noindent \textbf{Improve tool support for architecture related Technical Debt}\\
    As previously stated, code is only one of the artifacts from which Technical Debt can be identified. The code may be free of all quality issues but still not follow a well-reasoned architecture, demonstrating susceptibility to critical issues such as high-latency or security flaws, which are hard to detect and fix as the system grows in complexity. Identifying and managing Technical Debt at the architectural abstraction level often requires insight from humans and rich context understanding, which current tools minimally offer, as discussed in more detail in the role of architecture (Section \ref{architectureLandscape}). \\
    
     \noindent\textbf{Develop workflow-based Technical Debt Management Tools}\\
    A Technical Debt Management workflow represents different steps involved in identifying, validating, measuring, prioritizing, monitoring, reporting, and repaying TDIs. A Technical Debt Management workflow needs to be not only set up and configured appropriately but also integrated into organizations' practices and toolchains. Although existing tools support certain aspects of Technical Debt Management, they are fragmented and do not lend themselves to a consistently followed Technical Debt Management process.\\
    
    \noindent \textbf{Create diverse benchmarks for tools}\\
    Existing tools, even static analysis tools that detect code quality issues, do not report issues consistently. The current literature lacks detailed, precise, consistent and repeatable definitions of key concepts such as code quality metrics and manually validated benchmarks. Code quality benchmarks alone are insufficient. Comprehensive benchmarks should incorporate other development artifacts (\eg, architecture, build scripts, unit tests) and account for domain- and language-specific variations to offer guidance to fine-tune tools. \\
    
    \noindent\textbf{Clarify the influence of Artificial Intelligence}\\
    Software development tools are evolving to include features that incorporate AI-based features, including but not limited to auto-complete, refactoring recommendations, auto-generation of different code artifacts ranking from build scripts to unit tests, or ways to correct identified defects. There are a range of AI-based approaches that such new features rely on, ranging from search-based reasoning to the state-of-the-art generative AI. 
    AI-based features are likely to help software engineers avoid common repetitive mistakes. This development will help practitioners avoid misinterpreting basic code quality concerns as Technical Debt. However, AI-based approaches are probabilistic and non-deterministic, and they are prone to errors, some of which could evolve into Technical Debt. Therefore, future Technical Debt tools will need to account for both ends of this spectrum—reducing common quality errors and their misclassification as Technical Debt and accounting for AI-driven inaccuracies resulting in Technical Debt.

\subsection{Data Collection}

Data collection is critical for Technical Debt Management. Below, we illustrate state of the art (which metrics are collected, where data are obtained, how these metrics are used), and persistent challenges.

\subsubsection{State of the Art}


Data collection answers four generic questions: 
\begin{enumerate}
    \item Does Technical Debt even exist?,
    \item What is the value/cost of Technical Debt?,
    \item How is team velocity affected by Technical Debt?,
    \item Do developers perceive these issues as Technical Debt?
\end{enumerate} 

\noindent A diverse range of data supports these questions: code-level metrics (\eg, code smells, security issues, test coverage, Self-Admitted Technical Debt)~\cite{Avgeriou2021,Biazotto2024}, runtime information (\eg, performance hotspots, API usage patterns)~\cite{Besker2018,Jaspan2023}, documentation feedback (\eg, issues, Slack channels)~\cite{Tan2023}, and architecture analyses~\cite{Besker2018b}. 
A complete picture requires collecting developer experience and perceptions through surveys, interviews, and experience sampling to confirm whether identified artifacts represent Technical Debt~\cite{Ernst2015,Jaspan2023}. Technical Debt can be seen as the difference between "what the code is" and "what the code ought to be"; this latter piece requires assessment by asking developers directly.\\

\noindent\textbf{Measure Value and Cost}\\
Value and cost assessments currently consider both external and internal factors. For example, external value assessments of taking on Technical Debt might include customer impact via Net Promoter Score, the value in hitting the market earlier, defect rates, or the speed of closing customer-reported issues. Internal cost assessments might include developer satisfaction, friction in obtaining information, or speed of developing new features. These indicators shed light on the hidden costs of Technical Debt and the resources typically needed to mitigate it.\\

\noindent\textbf{Identify Technical Debt}\\
Existing approaches rely on multiple data sources to identify Technical Debt:
\begin{itemize}
\setlength\itemsep{0em}
\item \emph{Code data}: Security vulnerabilities, code smells, architectural degradation, and test coverage gaps~\cite{Avgeriou2021,Biazotto2024}.
\item \emph{Runtime data}: Execution traces reveal which components are most relevant, where time is spent, and how APIs are used~\cite{Besker2018,Jaspan2023}.
\item \emph{Documentation and Forums}: Developer feedback, issue trackers, user posts, and internal communication channels highlight missing or unclear documentation~\cite{Ernst2015,Tan2023}.
\item \emph{Architecture Issues}: Structural misalignments and the need for migrations or modernizations~\cite{Ernst2015,Besker2018b}.
\end{itemize}

\noindent Existing research and tools (\eg, static analyzers, issue trackers, DevOps dashboards) are helping us identify many forms of Technical Debt~\cite{Biazotto2024}. However, teams often need additional context to confirm severity and real-world impact of flagged issues~\cite{Jaspan2023}.\\

\noindent\textbf{Team Velocity and Tracking}\\
To estimate team velocity, teams routinely use metrics from version control, continuous integration, and issue tracking systems (\eg, cycle time, story points, deployment frequency)~\cite{Besker2018,Kruchten2012}. Flow-related metrics reveal potential stalls in work, indicating areas where Technical Debt may be impeding progress~\cite{Jaspan2023}. Missed milestones, frequent rollbacks, and prolonged rework cycles also serve as signals of underlying technical or architectural shortcomings~\cite{Ernst2015,Besker2018}.\\

\noindent\textbf{Developer Confirmation and Perception}\\
In practice, teams do attempt to supplement tool-based detection with active developer confirmation. They employ surveys, interviews, and rating exercises to validate whether items flagged by automated metrics align with day-to-day developer experiences~\cite{Besker2018b,Jaspan2023}. Self-admitted Technical Debt (\eg, comments, commit messages that label shortcuts) and community consensus (\eg, frequently referenced issues in trackers) further reflect the perceived severity and urgency of specific Technical Debt items~\cite{Tan2023,Jaspan2023}. Although these sources help us estimate the team's understanding, they do not always provide the full picture, prompting efforts to integrate developer perceptions more deeply into our processes.

%
\subsubsection{Challenges}
Improving how we benchmark Technical Debt data and applying new perspectives will significantly improve consistent treatment of Technical Debt. \\

\noindent\textbf{Benchmarks and Historical Context}\\
Despite widespread usage of multiple artifact types (code, issues, documentation, communication logs) with developer perspectives and cost estimates, we still see the \textbf{need for comprehensive benchmarks} to unify these data sources. Such benchmarks can help to categorize, compare, and refine Technical Debt Management practices and tools in a standardized way. However, designing benchmarks that are both representative and actionable remains elusive.\\


\noindent\textbf{Unaddressed Perspectives and Challenges}\\
Several \textbf{emerging factors are not yet fully integrated} into current workflows, such as coping with new regulations (such as new privacy rules or AI guidelines), managing AI-generated code, intangible aspects like trust and knowledge sharing, and legal/copyright issues (e.g., tracking who wrote which source code tokens). As a result, we continue to explore questions such as how to differentiate Technical Debt from general productivity issues, how to ensure representative and unbiased data collection, and how to correlate observable indicators with actual developer perceptions and long-term costs.\\


Ultimately, the existing data landscape is rich yet fragmented. While some code- and runtime-based metrics and flow indicators are commonly applied, combining them in practice is not trivial. Furthermore, variations in how tools measure Technical Debt make direct comparisons challenging. This variability complicates efforts to create universal benchmarks or reference points to guide systematic evaluations and best practices.


\subsection{The Role of Architecture}
\label{architectureLandscape}

We adopt the definition of \textbf{software architecture} as commonly accepted in research and practice: a set of design decisions~\cite{1620096}. 
The lack of focus on Technical Debt practice in architecture often relates to architecture-related challenges in practice. Within individual organizations, there is significant variation in the rigor and the methods applied across projects. This variation becomes even more pronounced when making broader observations across software engineering domains, \eg, IT, cloud and web applications, or safety-critical avionics software. 

The effort needed to design a well-thought-out software architecture grows superlinearly as systems grow in size and with overall complexity (especially competing non-functional requirements). 
Larger systems and teams typically develop a software architecture early in the software development process, but fail to keep that architecture documentation relevant across the development and operational life cycles. Often, legacy systems have incomplete, ill-defined, or even no software architecture documentation, making it hard to evolve them due to accumulated Technical Debt.

\subsubsection{State of the Art} 

\noindent \textbf{Processes and Practices to manage Architecture Technical Debt}\\
The wide variability in architecting across software organizations is reflected in the various approaches used to perform the core Technical Debt Management activities~\cite{Li2015}: document, identify, measure, prioritize, and prevent Technical Debt. 

\textit{Documentation} of Architecture Technical Debt often uses traceability between requirements and architecture, as well as architecture and business; both types of traceability support the localization of changes when identifying and paying back the debt. More importantly, the documentation of architecture (including architecture-significant requirements) either through Architecture Decision Records (ADRs)~\cite{keeling2017} or through architecture views and models \cite{Kruchten1995, Rozanski2011, bass2012} for more safety-, mission-, business-critical systems domains, can be crucial in capturing Technical Debt. For example, ADRs routinely capture tradeoffs between quality properties (including maintainability and evolvability) and sometimes even mention incurring Technical Debt as part of those tradeoffs. Violations of architecture decisions (\eg, bypassing defined interfaces in the source code, or improper use of security-induced designs) also often entail architecture-level Technical Debt.

\textit{Identification} of Architecture Technical Debt uses several practices that are not specific to Technical Debt Management. Automated testing (\eg, integration and regression testing) can form the basis for uncovering architecture issues resulting in Technical Debt. More importantly, code reviews, scenario-based architecture evaluations, peer reviews, deep dives (walkthroughs from subject matter experts), design reviews, and inspections of architecture models are common practices in uncovering Technical Debt from architecture. Existing tools have some support for assessing architectural aspects related to how the system is structured as these can be deemed partially from the code. For example, dependency checking tools (\eg, ArchUnit, Lattix) can support identifying layer violations or cyclic dependencies in the source code. 

Regarding \textit{measuring} Technical Debt from architecture, different teams use different metric sets based on different quality models. Often, both metric sets and quality models are customized for the particular organization and/or system. Since architecture documentation \textit{per se} is often missing, data for such Technical Debt measurement come from combining data from source code, issue trackers, and commit data. This results in metrics such as cohesion and coupling, code reuse, code duplication, logical coupling, and data modularity to dominate. Issue trackers may reveal that certain types of maintenance tasks consume an unexpectedly high amount of time or require many changes for simple functionality, thus potentially hinting at hidden Technical Debt.

\textit{Prioritization} of Architecture Technical Debt items in some organizations follows the standardized approach of including those items in the backlog alongside features, bugs, and infrastructure. It is important to note that some of these items may be related, need to be resolved together, or may hint at same issues, just assessed from different artifacts. Depending on the organization and team, prioritization may be based on various characteristics of Technical Debt Items such as their business value, technical effort, safety- or security-level attributes, root cause, and software development lifecycle phase where the item originated. Some organizations have included an architecture micro-cycle within agile meetings (\eg, the daily standup), where architecture decisions and their related Technical Debt are discussed and prioritized. However, prioritizing specific items considering their impact on the development process may require a deep understanding of the entire system and its planned evolution roadmap.

Finally, \textit{preventing} Technical Debt arising from architecture issues is the least supported and practiced activity~\cite{sas2022evolution}. Most organizations are not competent at prevention other than striving for reusable designs, \eg, through product line engineering. This is still an open research problem. Ideally, the same suite of tools should be integrated to support all aspects of Technical Debt (identify, document, measure, prioritize, and prevent) in a CI/CD pipeline to enable its continuous management. Educating architects and developers on best practices, patterns, and anti-patterns of Technical Debt for a particular development context or technology domain is essential. \\

\noindent \textbf{Tools to manage Architecture Technical Debt}\\
Similar to the broad spectrum of processes and practices utilized to manage Technical Debt in architecture, the tools to support analyzing it are also quite dispersed across industry and software engineering domains. In fact, and contrary to Technical Debt assessed from code, there are no industry-standard and definitive tools for identifying, measuring, documenting, prioritizing, or preventing architecture issues resulting in Technical Debt. 

Practitioners often use tools that serve other specialized purposes to perform some of these activities. For example, tools that perform static source code analysis for dependency analysis or uncover design-level smells are used for architecture-related Technical Debt identification. For the same activity, tools for conformance analysis (implemented vs. intended architecture) or data flow and control flow analysis are commonly used, although such tools also concentrate more on code-level than architecture-level issues. Finally, tools for software composition analysis allow practitioners to identify outdated versions of components and compliance with regulatory requirements, both of which indicate Architecture Technical Debt.

Large organizations that understand the importance of proactive Technical Debt Management, in addition to the aforementioned specialized tools, sometimes build \textit{bespoke} company-specific tools, and these tools can be tailored to integrate with the architecture documentation format. These customized tools best address architecture concerns as they can be tailored to address specific (software architecture) organizational needs. In contrast, extending commercially available tools is often deployed as an afterthought to meet an identified business need. 
However, an extension of an existing tool is typically constrained by the underlying tool implementation. 

\subsubsection{Challenges}
Industry use cases consistently expose the difficulties of managing architecture-related Technical Debt, for example, the consequences of poor design decisions, violation of best-practice architectural patterns, or suboptimal technology choices \cite{avgeriou_2023}. Architecture tradeoffs are more implicit and complex, and often are difficult to track in development artifacts to assess the incurred Technical Debt compared to others, such as code. In addition, architecture tradeoffs are made early in the software development process, and have far-reaching consequences, some of which result in expensive Technical Debt \cite{avgeriou_2023}. Furthermore, the rationale for architectural tradeoffs is elusive, often not directly visible within source code, and rarely documented within life-cycle artifacts. 
Consequently, while there is broad agreement on the criticality of architecture in managing Technical Debt, available tooling, process guidance, and best practices are far from ideal. To elevate the role of architecture in managing Technical Debt, progress needs to be made in the following areas. \\

\noindent \textbf{Focus on Standardized Architecture-level Tooling}\\
Software engineers and researchers tend to focus on easily identifiable debt (code) rather than the higher impact and more challenging debt (architecture). We established earlier that organizations either use commercial tools, which primarily target source code that are then customized especially for architecture, or they build organization-specific tools. 
Neither case is ideal. 
What industry really needs is a movement towards standardized analysis tools for architecture that can be seamlessly integrated with the developer ecosystem and CI/CD pipelines. For example, specific architecture rules can be embedded in dependency analysis tools to prohibit certain components from communicating with others and triggering alerts in the CI/CD process to prevent unintentional architecture violations and Technical Debt to creep in. 

Both commercial and research tools have their limitations. On the one hand, commercial tools need to become more transparent in how they identify and measure architecture and its relationship to Technical Debt. Software practitioners often struggle with the steep learning curve of these tools, and are challenged to customize them to achieve their business goals. On the other hand, emerging research tools allow for detecting self-admitted Technical Debt (currently not offered by commercial tools) but do not scale or perform well in architecture-level issues. Finally, neither commercial nor research tools currently measure interest (the cost of change because of essential architecture and related changes as a result of Technical Debt) despite being a decisive factor in prioritizing and repaying Technical Debt items.

The lack of tools specializing in architecture management leads to the scenario where software architects either don't manage their Technical Debt or may attempt to apply inadequate and/or ad hoc practices. Eventually, when architecture decisions and their realization are not sufficiently managed, the architecture and its documentation become disconnected from the implementation, and eventually turns into a ``paper architecture'', leading to further accumulating Technical Debt, which is further hidden in the architecture, a feedback cycle which makes Architecture Technical Debt harder to identify and assess.\\

\noindent \textbf {Identify Useful Architecture Metrics} \\
The metrics used as proxies (\eg, complexity and cohesion/coupling for larger code modules) often fail to provide practical insights, as their interpretations vary from their intended meanings; they are often not well understood by software practitioners, let alone trusted for making architecture decisions~\cite{sarkar2006api}. Such metrics need further refinement and context to interpret for reliable architecture measurement otherwise, they are easily dismissed as non-value adding. 

Furthermore, the structural metrics provide only one perspective about the system and its architecture, missing architecture concerns related to run-time and deployment aspects and the accumulating Technical Debt.  These structural metrics, even if they do not capture all architecture concerns to assess Technical Debt, also need to be integrated into quality models, as using single metrics is often misleading. It is also important to underscore that since different tools use different (and inconsistent) metrics, comparing tools across projects and companies is almost impossible. \\

\noindent \textbf {Expose Architecture Decisions}\\
While documenting architecture decisions in Architecture Decision Records (ADRs) is crucial for effective management, often ADRs are incomplete, inconsistent, or even completely missing. Again, tools, especially those integrated into CI/CD, can help enforce up-to-date and valid ADRs. AI-based tools are emerging that auto-generate low-level and high-level software requirements as well as decisions and designs from existing (Open-Source Software) code bases. However, auto-generated artifacts still need to be reviewed and updated, as necessary, to ensure correctness and completeness. Technical Debt persists until these human-based reviews have occurred and are dealt with, making accurate architecture assessment via automation difficult, resulting in the unexpected, hidden cost of accumulating Technical Debt because no one paid attention to the architecture. Architecture tooling must rapidly evolve to keep pace with these developments and support Technical Debt Management. As a bonus, focusing on architecture-level issues will further enable to resolve Technical Debt items on code, testing, and operations. 

\begin {comment}
\subsubsection{Challenges}
Industry use cases consistently expose the difficulties of managing architecture related Technical Debt, for example, the consequences of poor design decisions, violation of best-practice architectural patterns, or suboptimal technology choices \cite{avgeriou_2023}. Architecture tradeoffs are more implicit and complex, and often are difficult to track in development artifacts to assess the incurred Technical Debt compared to others, such as code. In addition, architecture tradeoffs are made early in the software development process, and have far-reaching consequences, some of which result in expensive Technical Debt \cite{avgeriou_2023}. Furthermore, the rationale for architectural tradeoffs is elusive, often not directly visible within source code, and rarely documented within life-cycle artifacts. 
Consequently, while there is broad agreement on the criticality of architecture in managing Technical Debt, available tooling, process guidance, and best practices are far from ideal. This may be a consequence of  ``lamp post'' research, as researchers tend to focus on easily identifiable debt (code) rather than the higher impact and more challenging debt (architecture).

We established earlier that organizations either use commercial tools, which primarily target source code customized especially for architecture, or build organization-specific tools. 
Neither case is ideal. 
What the industry seeks is a movement towards standardized architecture analysis tools that can be seamlessly integrated with the developer ecosystem and CI/CD pipelines. For example, specific architecture rules can be embedded in dependency analysis tools to prohibit certain components from communicating with others and trigger alerts in the CI/CD process to prevent unintentional architecture violations and Technical Debt from creeping in. 

Both commercial and research tools have limitations. On the one hand, commercial tools need to become more transparent in how they identify and measure architecture and its relationship to Technical Debt. Software practitioners often struggle with the steep learning curve of these tools and are challenged to customize them to achieve their business goals. On the other hand, emerging research tools allow for detecting self-admitted Technical Debt (currently not offered by commercial tools) but do not scale or perform well in architecture-level issues. Finally, neither commercial nor research tools currently measure interest (the cost of change because of essential architecture and related changes as a result of Technical Debt), despite being a decisive factor in prioritizing and repaying Technical Debt items.

The lack of tools specializing in architecture management leads to the scenario where software architects either don not manage their Technical Debt or may attempt to apply inadequate and/or ad hoc practices. Eventually, when architecture decisions and their realization are not sufficiently managed, the architecture and its documentation become disconnected from the implementation, and eventually turns into a ``paper architecture'', leading to further accumulating Technical Debt, which is further hidden in the architecture, a feedback cycle which makes Architecture Technical Debt harder to identify and assess.

The metrics used as proxies (\eg, complexity and cohesion/coupling for larger code modules) often fail to provide practical insights, as their interpretations vary from their intended meanings; they are often not well understood by software practitioners, let alone trusted for making architecture decisions~\cite{sarkar2006api}. Such metrics need further refinement and context to interpret for reliable architecture measurement otherwise, they are easily dismissed as non-value adding. 

Furthermore, the structural metrics provide only one perspective about the system and its architecture, missing architecture concerns related to run-time and deployment aspects, and the accumulating Technical Debt.  These structural metrics, even if they do not capture all architecture concerns to assess Technical Debt, also need to be integrated into quality models, as using single metrics is often misleading. It is also important to underscore that since different tools use different (and inconsistent) metrics, comparing tools across projects and companies is almost impossible. 

While documenting architecture decisions in ADRs is crucial for effective management, often ADRs are incomplete, inconsistent, or even completely missing. Again, tools, especially those integrated into CI/CD, can help enforce up-to-date and valid ADRs. AI-based tools are emerging that auto-generate low-level and high-level software requirements as well as decisions and designs from existing (Open-Source Software) code bases. However, auto-generated artifacts still need to be reviewed and updated, as necessary, to ensure correctness and completeness. Technical Debt persists until these human-based reviews have occurred and are dealt with, making accurate architecture assessment via automation difficult, resulting in the unexpected, hidden cost of accumulating Technical Debt because no one paid attention to the architecture. Architecture tooling must rapidly evolve to keep pace with these developments and support Technical Debt Management. As a bonus, focusing on architecture-level issues will further enable to resolve  Technical Debt items on code, testing, and operations. 
\end{comment}

\subsection{Socio-Technical Aspects}
The socio-technical aspects of Technical Debt refer to the interplay between social and technical factors influencing how Technical Debt is incurred, managed, addressed, and perceived by humans. 
In this section, we share a snapshot of 
existing research, tools and practices, and challenges for both research and tools/practices.  

\subsubsection{State of the Art}

Socio-technical aspects of Technical Debt encompass aspects such as communication and communication structures regarding Technical Debt, awareness of Technical Debt, decision-making regarding Technical Debt, changing the mindset of stakeholders or even company cultures regarding Technical Debt, and effects of Technical Debt perceived by humans.\\

\noindent \textbf{Research}
    In studying socio-technical aspects of Technical Debt, researchers largely rely on internal organizational research or studies~\cite{jaspan_defining_2023, McConnell2008a}.
    In contrast, researchers of socio-technical aspects rarely consider factors such as code quality and metrics, which are only used \eg, in gamification approaches~\cite{borg_industrial_2024} or as supporting factors to evaluate an intervention's outcome~\cite{crespo_carrot_2021}.
    
    Research methods used to study socio-technical aspects of Technical Debt typically comprise qualitative methods.
    For analyzing current Technical Debt Management approaches, researchers employ methods such as case studies~\cite{yli-huumo_sources_2014, Wiese2022}, interview studies~\cite{Verdecchia2020, Besker2019c, wiese_it_2023}, questionnaires (\eg, the InsighTD project)~\cite{InsighTD, Rios2018}, or observations~\cite{codabux2013managing}.
    They further used developer profiling \cite{codabux2020profiling, Besker2022}, \eg, to explore the influence of Technical Debt on developers' morale~\cite{Besker2019d, Besker2020d, Besker2022}.
    
    To influence social-technical aspects, researchers used action research studies to establish Technical Debt Management approaches ~\cite{borup_deliberative_2021, Detofeno_2021}, classroom studies for enhancing Technical Debt awareness~\cite{Tonin2017, crespo_carrot_2021}, and gamification~\cite{Crespo2022, borg_industrial_2024}, 
    and visualization approaches~\cite{Kaiser2011} to enable Technical Debt monitoring and, thus, promoting Technical Debt awareness in the long term. 
    We consider all methods valuable but acknowledge that their use does not guarantee results.\\
    
\noindent \textbf{Tools and Practices}\\
    As explained in Section \ref{sec:NextGenTooling}, there are currently a few tools that support Technical Debt Management; however, we did not find any tools that specifically incorporate socio-technical aspects. 
    We recognize that issue-tracking tools (\eg, Jira~\cite{Jira}) are flexible enough to support socio-technical aspects \eg,  by defining a specific Technical Debt issue type with attributes that might trigger discussions on Technical Debt issues. 

    There are many current practices for managing Technical Debt, which may provide opportunities for incorporating socio-technical aspects, especially bringing together multiple stakeholder perspectives. Examples from current work include:
    \begin{itemize}
        \item Using a Technical Debt backlog~\cite{Kruchten2012a} to promote discussions and conscious decision-making regarding Technical Debt incurrence and prioritize Technical Debt repayments.
        \item Establishing a champion for Technical Debt within the project or organization for enhancing Technical Debt awareness or a culture change~\cite{jaspan_defining_2023}. 
    \item Create and review metrics/dashboards/visualizations for management and prioritization~\cite{Kaiser2011, Mendes2019, Ozkaya2023RTC} to enhance Technical Debt awareness and support rational decision-making.
    \item Conducting and analyzing recurring surveys~\cite{jaspan_defining_2023} to communicate issues from developers to management stakeholders. 
    \item Incentivizing Technical Debt Management~\cite{Besker2020d} to encourage a change of mindset regarding Technical Debt, practices such as employing strategic code reviews ~\cite{freire_surveying_2020, Yli-Huumo2016}, or performing bug bashes/hackathons~\cite{codabux2014quality}.
    \item Establish a structured budget to address Technical Debt~\cite{McConnell2008a} to further contribute to incorporating multiple stakeholder perspectives (e.g., technical and organizational) to achieve socio-technical congruence.
\end{itemize}

\subsubsection{Challenges}
    \textbf{Bridge Socio-Technical Gaps with Quantitative and Qualitative Insights}
    
    \noindent Technical Debt Management encompasses complex social dynamics influencing stakeholder decision-making, collaboration, and mindsets. 
    A major challenge is the reliance on studies with relatively small sample sizes to draw conclusions, which compromises scientific rigor and generalizability of the results. 
    However, a quantitative evaluation of socio-technical aspects resulting from Technical Debt Management initiatives, such as awareness or mindset changes, is complicated. 
    Thus, research opportunities include creating new methods and metrics for evaluating socio-technical aspects or adopting them from other research domains \eg, awareness research in psychology or research on company cultures from the business domain.\\
    
    

    \noindent \textbf{Integrating Socio-Technical Features into Existing Toolsets}
    
    \noindent Another key challenge affecting the socio-technical dimension of Technical Debt is the absence of dedicated Technical Debt-specific features within an organization's existing toolsets. For example, adding a reminder within backlog items to nudge the developers to discuss potential Technical Debt might help prevent Technical Debt. Furthermore, enhancing issue tracking systems with a dedicated backlog item type for Technical Debt, incorporating attributes to capture both principal (implementation cost) and interest (ongoing maintenance cost), could facilitate more informed decision-making. These enhancements should be integrated into current toolsets rather than introducing new tools, thereby increasing stakeholder acceptance and minimizing disruption to established processes.\\


    \noindent Ongoing challenges in Technical Debt Management include the absence or inefficacy of practices for handling high-cost Technical Debt items, Technical Debt prioritization methods, and the inclusion of the right stakeholders (\eg, business people). \\

      \noindent \textbf{Promote Management Support and Cultural Shifts for Proactive Technical Debt Management}
      
      \noindent  An opportunity for improvement lies in fostering a shift in upper management's perspective on Technical Debt, leading to increased support for Technical Debt Management initiatives.
  With increased organizational commitment and a corresponding cultural shift toward proactive Technical Debt Management, practices such as organizational training programs on Technical Debt, as well as project- or personnel-based incentives designed to systematically reduce Technical Debt while fostering developer engagement and morale, could be implemented.
    Additionally, establishing Technical Debt champions at different organizational levels could serve as a mechanism to advocate for Technical Debt awareness, incorporating socio-technical perspectives, improve its visibility, and promote best practices. The formation of a Technical Debt Community of Practice, where these champions collaborate to share knowledge, standardize management approaches, and drive organizational alignment in Technical Debt remediation efforts, can contribute to Technical Debt Management.
    


\section{Roadmap}
\label{sec:roadmap}
Incorporation of the principles formulated by this manifesto into routine software development practices can significantly improve Technical Debt Management. To facilitate the dissemination, implementation, and continuous development of the principles, we formulate a five-phase roadmap, to be used as a yardstick to measure progress towards the manifesto's realization. 
The phases may interleave; for example, selected Phase 2 milestones may be reached before completing Phase 1 milestones. 

The phases serve as a coarse-grained orientation to enable progress in a 3-5 year time frame, not as a strict separation of work items. The milestones recommended in this roadmap are aspirational to help track change and progress. We anticipate that the community will contribute to making progress on all fronts simultaneously and share lessons learned and results as progress is achieved.  

\subsection{Phase 1: Manifesto Dissemination and Community Engagement}
Upon manifesto release, the authors shall share it at renowned software engineering venues (\eg, the International Conference on Software Engineering, TechDebt conference, IEEE Software Magazine, Developer Productivity Engineering Conference, Agile Conference, QCON, GOTO, KubeCon, PyCon). Tool vendors will be encouraged to distribute the one-pager (with the Values, Beliefs and Principles) to their customers and further disseminate it via social media, podcasts, and newsletters, to reach a wide audience. Practitioners will test drive the manifesto principles and provide lessons learned, e.g through conference presentations or short YouTube training videos, while researchers will prepare technical briefings on selected manifesto principles to further support dissemination.

Manifesto feedback shall be collected from various channels from different stakeholders, including conferences, open-source communities, and social media. This can include ideas on how to implement manifesto principles, as well as change requests or refinements for future versions. Those working on Technical Debt tooling and research will be encouraged to establish collaborations with related fields (\eg, refactoring, mining software repositories, software evolution, software architecture, software economics, developer productivity) or on selected manifesto principles. We especially encourage potential collaborations with other disciplines (\eg, economics, social sciences, psychology, systems engineering). 

\subsubsection{Milestones Phase 1}
\begin{enumerate}
    \item Public launch of the manifesto.
    \item Document summarizing feedback finalized twelve months after public launch.
    \item Concrete collaborations with related research fields (joint research and organizational initiatives) are established by the community.
    \item Concrete collaborations with software development communities are established (such as users of the Scaled Agile Framework) to incorporate minimum training and practices. 
\end{enumerate}

\subsection{Phase 2: Encourage Manifesto Principle Transition into Practice}
While practitioners can directly apply the principles formulated within the manifesto, the software engineering community and organizations that publish practices and guidelines can develop guidelines, processes, and tools to further support and facilitate the realization of those principles in practice. This is the main focus of Phase 2, which comprises three parts.

The first part entails codifying and communicating processes and policies around Technical Debt Management. To foster shared responsibility (\ie, \principle{P6 - Visible Technical Debt} and \principle{P9 - Conscious Tradeoff Decisions}) appointing developers and software architects into ``Technical Debt champion'' roles can be beneficial. Best practices for this role shall be collected and published. A minimum viable Technical Debt Management process shall be formulated as well as more sophisticated processes, thus addressing \principle{P4 - no one-size-fits-all metric} as well as \principle{P5 - seamless and integrated toolchains}. The signatories of this manifesto commit to collecting and sharing their experiences as relevant. Standard questionnaires for collecting qualitative and subjective data on Technical Debt shall be published and support \principle{P3 - comprehensive data collection}. These efforts will result in guidelines for appropriately sizing Technical Debt Management efforts within a given context.

The second part of Phase 2 concerns tool vendors and researchers working on next-generation Technical Debt Management software tools. In the short term, understanding whether existing established metrics and standards, such as DORA~\footnote{https://dora.dev/guides/dora-metrics-four-keys/} and Flow metrics~\footnote{https://devops.com/measuring-value-streams-by-analyzing-flow-metrics/}, support Technical Debt Management can assist in avoiding single Technical Debt metric tendencies. Short-term enhancements could rely on AI to integrate ``fuzzy'' contextual information into metric-centered tools and improve the interpretability of their results, which fosters  \principle{P2 - alignment with context}. AI may also be instrumental in tailoring default metric sets within Technical Debt Management tools to a given project situation, thus addressing \principle{P4 - no one-size-fits-all metric}. These are ideas researchers can take on but can also be shared by organizations using any of these tools and industry best practices and standards.   

Furthermore, tool development will likely require novel approaches to conform to \principle{P7 - elevated role of architecture in Technical Debt Management}. Specifically,  software tools may need to consider architecture decision records, diagrams, design (anti-)patterns, or other forms of codifying architectural knowledge. This will need to go significantly beyond the classical focus on coupling and cohesion metrics. The tools must support developers and architects to follow easily \principle{P9 - making conscious and informed tradeoff decisions}, instead of simply presenting a generic source code analysis and leaving users alone in interpretation and decision making. Interoperability standards between tools and tighter integration with issue tracking systems (\eg, populating Technical Debt backlog item lists) could enhance Technical Debt Management tools.

The third part of Phase 2 is meant to prepare for the validation of novel tools later in Phase 3; to this end, researchers shall already work on defining one or multiple community-wide benchmarks, supported by practitioners and tool vendors to realize \principle{P8 - Technical Debt benchmarks}. Such benchmarks shall be openly available and cover a range of software engineering artifacts, including code, requirements, and design rationale. Tool vendors shall release exemplar benchmarks along with the features their tools offer. An open-source code base with best-practice violations is not sufficient for a real Technical Debt benchmark. The benchmark would also need to provide a list of manually crafted ``real'' Technical Debt items, as defined by subject matter experts. This includes their description with their assumed extent, estimated impact, and possibly plans to fix them. The benchmark would also need to provide a prioritized feature backlog for the planned evolution of the system, so that the Technical Debt items can be put into context. Various tests shall be performed using such a benchmark, including finding out if the hand-crafted Technical Debt items can be automatically generated by an analysis of the available artifacts, excluding the handcrafted Technical Debt item list.

\subsubsection{Milestones Phase 2}
\begin{enumerate}
    \item Shared ``standard'' best-practice processes and guidelines for Technical Debt Management.
    \item Agreement on a minimum viable Technical Debt Management process.
    \item Update of at least three commercial tools with special features targeting the manifesto principles, such as establishing a dedicated Technical Debt label in default configurations.
    \item Launch of a community-wide Technical Debt benchmark (version 1.0) for at least one programming language.
\end{enumerate}

\subsection{Phase 3: Validation of New Processes and Tools}
Phase 3 may be interleaved with Phase 2 and focus on validating best practices, processes, and novel tools according to the manifesto principles. Practitioners in software development companies shall support researchers in carrying out empirical research on Technical Debt Management. Case studies on Technical Debt Management practices according to the new principles shall include quantitative and qualitative methods. Such studies would highlight short-term benefits as well as long-term improvements or negative side effects. 

Longitudinal studies on Technical Debt Management spanning several years are still rare, since they require sustained effort from multiple parties and may last longer than typical funded research projects or PhD work. However, principle \principle{P5 - Seamless and integrated Technical Debt Management toolchains} could lead to highly automated evaluation approaches that may require only small efforts for sustained data collection.

Tool vendors and researchers shall utilize the Technical Debt benchmarks developed during Phase 2 to test and improve their approaches. A leaderboard and yearly tool competitions on focused aspects of the benchmark shall be established. Besides immediate feedback on the capabilities of new methods and tools, the benchmark results will also provide insights for improving the benchmarks themselves in Phase 5. If tools provide a high degree of automation for Technical Debt identification and Technical Debt Management, more sophisticated benchmarks for even more challenging situations could be designed. This includes testing AI agents to automatically refactor or re-design the system to fix the listed Technical Debt items.

In addition to such project-specific activities, further empirical research shall be conducted on the adherence to the manifesto principles in the commercial software market. Surveys and interviews will provide feedback on the best practices and input for refining and extending the principles. 

\subsubsection{Milestone Phase 3}
\begin{enumerate}
    \item Ideally, at least three publicly reported case studies with commercial systems, whose development teams applied the principles and new processes for at least 6 months.
    \item First tool competition with the Technical Debt benchmark held at the TechDebt conference, where efficiency in different tasks (e.g. identifying, measuring, prioritizing, resolving
    ) is assessed against the community-wide Technical Debt benchmark of Phase 2.
    \item At least 25\% of software development teams report on having a dedicated role ``Technical Debt champion'' in a representative survey.
\end{enumerate}

\subsection{Phase 4: Institutionalization and Scaling}
After tool vendors and researchers have validated novel methods and tools for Technical Debt Management in Phase 3, the roadmap now foresees a phase of institutionalization and scaling of the Technical Debt Management principles. 

The principles shall be refined and enhanced and subsequently institutionalized by integrating them into established bodies of knowledge, \eg, the ACM Computer Science curriculum. 
This could enable widespread adoption as well as support accreditation of training programs. The TechDebt community shall release recurring awards and create recognition programs for teams and individuals who provide an exemplary implementation of Technical Debt Management practices. A Technical Debt Management ``hall-of-fame'' website shall be created, sharing success stories and case studies spanning companies and universities. 

Scaling the principles refers to reaching even more development teams and potentially using AI tools to automate Technical Debt Management tasks. At the time of writing, the impact of novel AI tools on the practice of Technical Debt Management is hard to predict. AI tools could potentially autonomously populate Technical Debt backlogs based on newly committed source code, thus further automating \principle{P6 - making Technical Debt visible}. It is conceivable that a substantial portion of refactoring activities will also be carried out autonomously by AI-agents in many projects, thus reducing the manual effort for paying down Technical Debt interest.  

\subsubsection{Milestones Phase 4}
\begin{enumerate}
    \item Technical Debt Management principles integrated into SWEBOK or equivalent software engineering curriculum guides.
    \item Technical Debt Management ``hall-of-fame'' created and populated with at least five success stories.
    \item A taxonomy of the different types of Technical Debt (code, design, test debt etc.) has been established and validated by practitioners and tool vendors.
    \item At least 50\% of software development teams report on having a dedicated role ``Technical Debt champion'' in a representative survey.
    \item Lessons learned on the current capabilities and essential limitations of AI tools for Technical Debt Management are shared.
    \item At least 30\% of software development teams report on emphasizing the role of software architecture in Technical Debt Management.
\end{enumerate}

\subsection{Phase 5: Continuous Improvement}
As the understanding of Technical Debt Management improves with novel methods, tools and data, the values, beliefs, and principles stated in the manifesto may need a revision or extensions. Phase 5 shall be interleaved with Phase 3 and 4 and focus on continuous refinement for the practice of Technical Debt Management itself.

Regular evaluations shall be performed to capture the state-of-practice, combined with evidence from empirical studies by organizations as well as at conferences. More standardized evaluation baselines and metrics shall be defined, so that progress or regressions on the state-of-practice can be made visible. Maturity models and automated compliance checks shall support improving practices even further.

\subsubsection{Milestones Phase 5}
\begin{enumerate}
    \item Impact analysis on how the manifesto influenced Technical Debt Management practices published.
    \item At least 80\% of software development teams report on having a dedicated role ``Technical Debt champion'' in a representative survey and apply commonly agreed upon Technical Debt Management practices.
\end{enumerate}


\section{Signatories}
The manifesto is endorsed by the participants of the Dagstuhl Perspectives Workshop 24452 on \textit{Reframing Technical Debt}:\\

\participant Apostolos Ampatzoglou, University of Macedonia, GR
\participant Paris Avgeriou, University of Groningen, NL
\participant Lodewijk Bergmans, Software Improvement Group, NL
\participant Markus Borg, CodeScene - Malmö, SE 
\participant Alexandros Chatzigeorgiou, University of Macedonia, GR
\participant Marcus Ciolkowski, QAware, DE
\participant Zadia Codabux, University of Saskatchewan, CA 
\participant Stefano Dalla Palma, Adyen, NL
\participant Florian Deißenböck, CQSE, DE 
\participant Philippe-Emmanuel Douziech, CAST, DE
\participant Neil Ernst, University of Victoria, CA 
\participant Daniel Feitosa, University of Groningen, NL 
\participant Michael Felderer, DLR, DE 
\participant Collin Green, Google, US 
\participant Ciera Jaspan, Google, US 
\participant Ron Koontz, The Boeing Company, US
\participant Heiko Koziolek, ABB, DE 
\participant Christof Momm, SAP, DE
\participant Brigid O'Hearn, Carnegie Mellon University, US
\participant Ipek Ozkaya, Carnegie Mellon University, US 
\participant Klaus Schmid, Universität Hildesheim, DE 
\participant Carolyn Seaman, University of Maryland, US 
\participant Tushar Sharma, Dalhousie University, CA
\participant Guilherme Horta Travassos, UFRJ / COPPE, BR 
\participant Roberto Verdecchia, University of Florence, IT
\participant Marion Wiese, Universität Hamburg, DE

\section{References}
\bibliography{ref}


\end{document}

%% file: bvpone.tex
    
\centertitle{Managing Technical Debt is essential to delivering high-quality software systems that are timely, are on-budget, and meet their stakeholder needs.}  \vspace{6pt}

\centertitle{To drive an effective Technical Debt Management practice}

\centertitle{We value:}
\begin{center}
    
Psychological safety and trust between technical and business stakeholders. 

Simple, actionable, value-based communication of Technical Debt to all stakeholders over excessive, minute, overwhelming details. 

Transparency, explainability, and replicability in the identification, measurement, and prioritization of Technical Debt. 

Both objective and subjective data collection and research methods to identify, measure, and prioritize Technical Debt over single-method approaches.

Architecture understanding across the team.
\end{center}

\centertitle{We believe that:}
\begin{center}
    
Sustainable software delivery requires proactive and continuous Technical Debt Management. 

The most effective Technical Debt Management is as automatic as possible and as manual as needed.

Technical Debt must be managed irrespective of how the software system is created, including generation of code and other artifacts by AI-assisted tools.

Items that developers observe as Technical Debt should be
addressed, even when they are not supported by metrics.

Not all all issues identified by stakeholders or tools are Technical Debt.
 
\end{center}

\centertitle{And we adhere to the following principles:}

\begin{center}
    
Share Responsibility for Technical Debt Management.

Manage Technical Debt in Alignment with its Context.

Collect Comprehensive Data for Technical Debt Management.

Avoid a One-Size-Fits-All Technical Debt Metric.

Build Seamless and Integrated Technical Debt Management Toolchains with Human Oversight.

Make Technical Debt Visible.

Elevate the Role of Architecture in Technical Debt Management.

Develop Fit-For-Purpose Technical Debt Benchmarks.

Make Intentional Technical Debt Trade-off Decisions.

\end{center}

%% file: main.bbl
\begin{thebibliography}{10}

\bibitem{alves2016identification}
Nicolli~SR Alves, Thiago~S Mendes, Manoel~G De~Mendon{\c{c}}a, Rodrigo~O Sp{\'\i}nola, Forrest Shull, and Carolyn Seaman.
\newblock Identification and management of technical debt: A systematic mapping study.
\newblock {\em Information and Software Technology}, 70:100--121, 2016.

\bibitem{DagstuhlAvgeriou2016}
Paris Avgeriou, Philippe Kruchten, Ipek Ozkaya, and Carolyn~B. Seaman.
\newblock Managing technical debt in software engineering (dagstuhl seminar 16162).
\newblock {\em Dagstuhl Reports}, 6(4):110--138, 2016.
\newblock \href {https://doi.org/10.4230/DagRep.6.4.110} {\path{doi:10.4230/DagRep.6.4.110}}.

\bibitem{avgeriou_2023}
Paris Avgeriou, Ipek Ozkaya, Alexander Chatzigeorgiou, Marcus Ciolkowski, Neil~A. Ernst, Ronald~J. Koontz, Eltjo Poort, and Forrest Shull.
\newblock Technical {Debt} {Management}: {The} {Road} {Ahead} for {Successful} {Software} {Delivery}.
\newblock In {\em 2023 {IEEE}/{ACM} {International} {Conference} on {Software} {Engineering}: {Future} of {Software} {Engineering} ({ICSE}-{FoSE})}, pages 15--30, May 2023.
\newblock URL: \url{https://ieeexplore.ieee.org/document/10449672}, \href {https://doi.org/10.1109/ICSE-FoSE59343.2023.00007} {\path{doi:10.1109/ICSE-FoSE59343.2023.00007}}.

\bibitem{Avgeriou2021}
Paris~C. Avgeriou, Davide Taibi, Apostolos Ampatzoglou, Francesca Arcelli~Fontana, Terese Besker, Alexander Chatzigeorgiou, Valentina Lenarduzzi, Antonio Martini, Athanasia Moschou, Ilaria Pigazzini, Nyyti Saarimaki, Darius~Daniel Sas, Saulo~Soares de~Toledo, and Angeliki~Agathi Tsintzira.
\newblock An overview and comparison of technical debt measurement tools.
\newblock {\em IEEE Software}, 38(3):61–71, 2021.
\newblock URL: \url{http://dx.doi.org/10.1109/ms.2020.3024958}, \href {https://doi.org/10.1109/ms.2020.3024958} {\path{doi:10.1109/ms.2020.3024958}}.

\bibitem{Avgeriou2020}
Paris~C. Avgeriou, Davide Taibi, Apostolos Ampatzoglou, Francesca Arcelli~Fontana, Terese Besker, Alexandros Chatzigeorgiou, Valentina Lenarduzzi, Antonio Martini, Nasia Moschou, Ilaria Pigazzini, Nyyti Saarimaki, Darius~Daniel Sas, Saulo~Soares de~Toledo, and Angeliki~Agathi Tsintzira.
\newblock An {Overview} and {Comparison} of {Technical} {Debt} {Measurement} {Tools}.
\newblock {\em IEEE Software}, 38(January), 2020.
\newblock \href {https://doi.org/10.1109/MS.2020.3024958} {\path{doi:10.1109/MS.2020.3024958}}.

\bibitem{bass2012}
Len Bass, Paul Clements, and Rick Kazman.
\newblock {\em Software Architecture in Practice: Software Architect Practice\_c3}.
\newblock Addison-Wesley, 2012.

\bibitem{Besker2018b}
Terese Besker, Antonio Martini, and Jan Bosch.
\newblock Managing architectural technical debt: A unified model and systematic literature review.
\newblock {\em Journal of Systems and Software}, 135:1–16, 2018.
\newblock URL: \url{http://dx.doi.org/10.1016/j.jss.2017.09.025}, \href {https://doi.org/10.1016/j.jss.2017.09.025} {\path{doi:10.1016/j.jss.2017.09.025}}.

\bibitem{Besker2018}
Terese Besker, Antonio Martini, and Jan Bosch.
\newblock Technical debt cripples software developer productivity: a longitudinal study on developers’ daily software development work.
\newblock In {\em Proceedings of the 2018 International Conference on Technical Debt}, page 105–114, Gothenburg, Sweden, 2018. ACM.
\newblock URL: \url{http://dx.doi.org/10.1145/3194164.3194178}, \href {https://doi.org/10.1145/3194164.3194178} {\path{doi:10.1145/3194164.3194178}}.

\bibitem{Besker2019d}
Terese Besker, Antonio Martini, and Jan Bosch.
\newblock {Software developer productivity loss due to technical debt—A replication and extension study examining developers' development work}.
\newblock {\em Journal of Systems and Software}, 156:41--61, 2019.

\bibitem{Besker2019c}
Terese Besker, Antonio Martini, and Jan Bosch.
\newblock {Technical Debt Triage in Backlog Management}.
\newblock In {\em 2019 IEEE/ACM International Conference on Technical Debt (TechDebt)}, TechDebt '19, pages 13--22. IEEE Press, 2019.

\bibitem{Besker2020d}
Terese Besker, Antonio Martini, and Jan Bosch.
\newblock {Carrot and Stick Approaches When Managing Technical Debt}.
\newblock In {\em Proceedings of the 3rd International Conference on Technical Debt}, TechDebt '20, pages 21--30, New York, NY, USA, 2020. Association for Computing Machinery.
\newblock \href {https://doi.org/10.1145/3387906.3388619} {\path{doi:10.1145/3387906.3388619}}.

\bibitem{Besker2022}
Terese Besker, Antonio Martini, and Jan Bosch.
\newblock {The use of incentives to promote technical debt management}.
\newblock {\em Information and Software Technology}, 142:106740, 2022.

\bibitem{Biazotto2024}
João~Paulo Biazotto, Daniel Feitosa, Paris Avgeriou, and Elisa~Yumi Nakagawa.
\newblock Technical debt management automation: State of the art and future perspectives.
\newblock {\em Information and Software Technology}, 167:107375, March 2024.
\newblock URL: \url{http://dx.doi.org/10.1016/j.infsof.2023.107375}, \href {https://doi.org/10.1016/j.infsof.2023.107375} {\path{doi:10.1016/j.infsof.2023.107375}}.

\bibitem{borg_industrial_2024}
Markus Borg, Amogha Udayakumar, and Adam Tornhill.
\newblock Industrial {Code} {Quality} {Benchmarks}: {Toward} {Gamification} of {Software} {Maintainability}, December 2024.
\newblock arXiv:2412.06307 [cs].
\newblock URL: \url{http://arxiv.org/abs/2412.06307}, \href {https://doi.org/10.48550/arXiv.2412.06307} {\path{doi:10.48550/arXiv.2412.06307}}.

\bibitem{borup2021deliberative}
Nichlas~B{\o}dker Borup, Ann Louise~Jul Christiansen, Sabine~H{\o}rdum Tovgaard, and John~Stouby Persson.
\newblock Deliberative technical debt management: An action research study.
\newblock In {\em Software Business: 12th International Conference, ICSOB 2021, Drammen, Norway, December 2--3, 2021, Proceedings 12}, pages 50--65. Springer, 2021.

\bibitem{borup_deliberative_2021}
Nichlas~Bødker Borup, Ann Louise~Jul Christiansen, Sabine~Hørdum Tovgaard, and John~Stouby Persson.
\newblock Deliberative {Technical} {Debt} {Management}: {An} {Action} {Research} {Study}.
\newblock In Xiaofeng Wang, Antonio Martini, Anh Nguyen-Duc, and Viktoria Stray, editors, {\em Software {Business}}, Lecture {Notes} in {Business} {Information} {Processing}, Cham, 2021. Springer International Publishing.
\newblock \href {https://doi.org/10.1007/978-3-030-91983-2_5} {\path{doi:10.1007/978-3-030-91983-2_5}}.

\bibitem{Cast}
{CAST Software}.
\newblock URL: \url{https://www.castsoftware.com/}.

\bibitem{codabux2020profiling}
Zadia Codabux and Christopher Dutchyn.
\newblock Profiling developers through the lens of technical debt.
\newblock In {\em Proceedings of the 14th ACM/IEEE International Symposium on Empirical Software Engineering and Measurement (ESEM)}, pages 1--6, 2020.

\bibitem{codabux2013managing}
Zadia Codabux and Byron Williams.
\newblock Managing technical debt: An industrial case study.
\newblock In {\em 2013 4th International Workshop on Managing Technical Debt (MTD)}, pages 8--15. IEEE, 2013.

\bibitem{codabux2014quality}
Zadia Codabux, Byron~J Williams, and Nan Niu.
\newblock A quality assurance approach to technical debt.
\newblock In {\em Proceedings of the International Conference on Software Engineering Research and Practice (SERP)}, page~1, 2014.

\bibitem{CodeScene}
{CodeScene}.
\newblock URL: \url{https://codescene.com/}.

\bibitem{crespo_carrot_2021}
Yania Crespo, Arturo Gonzalez-Escribano, and Mario Piattini.
\newblock Carrot and {Stick} approaches revisited when managing {Technical} {Debt} in an educational context.
\newblock 2021.
\newblock URL: \url{http://arxiv.org/abs/2104.08993}, \href {https://doi.org/10.1109/TechDebt52882.2021.00020} {\path{doi:10.1109/TechDebt52882.2021.00020}}.

\bibitem{Crespo2022}
Yania Crespo, Carlos López-Nozal, Raúl Marticorena-Sánchez, Margarita Gonzalo-Tasis, and Mario Piattini.
\newblock The role of awareness and gamification on technical debt management.
\newblock {\em Information and Software Technology}, 150(April):106946, 2022.
\newblock Publisher: Elsevier B.V.
\newblock \href {https://doi.org/10.1016/j.infsof.2022.106946} {\path{doi:10.1016/j.infsof.2022.106946}}.

\bibitem{Designite}
{Designite}.
\newblock URL: \url{https://www.designite-tools.com}.

\bibitem{Detofeno_2021}
Thober Detofeno, Sheila Reinehr, and Malucelli Andreia.
\newblock Technical {Debt} {Guild}: {When} {Experience} and {Engagement} {Improve} {Technical} {Debt} {Management}.
\newblock In {\em Proceedings of the {XX} {Brazilian} {Symposium} on {Software} {Quality}}, New York, NY, USA, 2021. Association for Computing Machinery.
\newblock \href {https://doi.org/10.1145/3493244.3493271} {\path{doi:10.1145/3493244.3493271}}.

\bibitem{doi:10.2307/2666999}
Amy Edmondson.
\newblock Psychological safety and learning behavior in work teams.
\newblock {\em Administrative Science Quarterly}, 44(2):350--383, 1999.
\newblock URL: \url{https://journals.sagepub.com/doi/abs/10.2307/2666999}, \href {https://arxiv.org/abs/https://journals.sagepub.com/doi/pdf/10.2307/2666999} {\path{arXiv:https://journals.sagepub.com/doi/pdf/10.2307/2666999}}, \href {https://doi.org/10.2307/2666999} {\path{doi:10.2307/2666999}}.

\bibitem{Ernst2015}
Neil~A. Ernst, Stephany Bellomo, Ipek Ozkaya, Robert~L. Nord, and Ian Gorton.
\newblock Measure it? manage it? ignore it? software practitioners and technical debt.
\newblock In {\em Proceedings of the 2015 10th Joint Meeting on Foundations of Software Engineering}, ESEC/FSE’15, page 50–60, Bergamo, Italy, 2015. ACM.
\newblock URL: \url{http://dx.doi.org/10.1145/2786805.2786848}, \href {https://doi.org/10.1145/2786805.2786848} {\path{doi:10.1145/2786805.2786848}}.

\bibitem{Fowler2009}
Martin Fowler.
\newblock {Technical Debt Quadrant}, 2009.
\newblock URL: \url{https://martinfowler.com/bliki/TechnicalDebtQuadrant.html}.

\bibitem{freire_surveying_2020}
Sávio Freire, Nicolli Rios, Boris Gutierrez, Darío Torres, Manoel Mendonça, Clemente Izurieta, Carolyn Seaman, and Rodrigo~O Spínola.
\newblock Surveying {Software} {Practitioners} on {Technical} {Debt} {Payment} {Practices} and {Reasons} for not {Paying} off {Debt} {Items}.
\newblock In {\em {ACM} {International} {Conference} {Proceeding} {Series}}, volume~10, 2020.
\newblock \href {https://doi.org/10.1145/3383219.3383241} {\path{doi:10.1145/3383219.3383241}}.

\bibitem{InsighTD}
{InsighTD Project – InsighTD Project}.
\newblock URL: \url{http://www.td-survey.com/}.

\bibitem{1620096}
A.~Jansen and J.~Bosch.
\newblock Software architecture as a set of architectural design decisions.
\newblock In {\em 5th Working IEEE/IFIP Conference on Software Architecture (WICSA'05)}, pages 109--120, 2005.
\newblock \href {https://doi.org/10.1109/WICSA.2005.61} {\path{doi:10.1109/WICSA.2005.61}}.

\bibitem{Jaspan2023}
Ciera Jaspan and Collin Green.
\newblock Defining, measuring, and managing technical debt.
\newblock {\em IEEE Software}, 40(3):15–19, May 2023.
\newblock URL: \url{http://dx.doi.org/10.1109/ms.2023.3242137}, \href {https://doi.org/10.1109/ms.2023.3242137} {\path{doi:10.1109/ms.2023.3242137}}.

\bibitem{jaspan_defining_2023}
Ciera Jaspan and Collin Green.
\newblock Defining, {Measuring}, and {Managing} {Technical} {Debt}.
\newblock {\em IEEE Software}, 40(3), May 2023.
\newblock Conference Name: IEEE Software.
\newblock URL: \url{https://ieeexplore.ieee.org/document/10109339/?arnumber=10109339}, \href {https://doi.org/10.1109/MS.2023.3242137} {\path{doi:10.1109/MS.2023.3242137}}.

\bibitem{Jira}
{Jira | Issue {\&} Project Tracking Software | Atlassian}.
\newblock URL: \url{https://www.atlassian.com/software/jira}.

\bibitem{jones2007investments}
Charles~P Jones.
\newblock {\em Investments: analysis and management}.
\newblock John Wiley \& Sons, 2007.

\bibitem{Kaiser2011}
Michael Kaiser and Guy Royse.
\newblock Selling the {Investment} to {Pay} {Down} {Technical} {Debt}: {The} {Code} {Christmas} {Tree}.
\newblock In {\em 2011 {Agile} {Conference}}, August 2011.
\newblock \href {https://doi.org/10.1109/AGILE.2011.50} {\path{doi:10.1109/AGILE.2011.50}}.

\bibitem{KASHIWA2022106855}
Yutaro Kashiwa, Ryoma Nishikawa, Yasutaka Kamei, Masanari Kondo, Emad Shihab, Ryosuke Sato, and Naoyasu Ubayashi.
\newblock An empirical study on self-admitted technical debt in modern code review.
\newblock {\em Information and Software Technology}, 146:106855, 2022.
\newblock URL: \url{https://www.sciencedirect.com/science/article/pii/S0950584922000258}, \href {https://doi.org/10.1016/j.infsof.2022.106855} {\path{doi:10.1016/j.infsof.2022.106855}}.

\bibitem{keeling2017}
Michael Keeling.
\newblock {\em Design It!: From Programmer to Software Architect}.
\newblock The Pragmatic Bookshelf, 2017.

\bibitem{Kruchten1995}
P.B. Kruchten.
\newblock The 4+1 view model of architecture.
\newblock {\em IEEE Software}, 12(6):42--50, 1995.
\newblock \href {https://doi.org/10.1109/52.469759} {\path{doi:10.1109/52.469759}}.

\bibitem{kruchten2019managing}
Philippe Kruchten, Robert Nord, and Ipek Ozkaya.
\newblock {\em Managing Technical Debt: Reducing Friction in Software Development}.
\newblock Addison-Wesley, 2019.

\bibitem{Kruchten2012}
Philippe Kruchten, Robert~L. Nord, and Ipek Ozkaya.
\newblock Technical debt: From metaphor to theory and practice.
\newblock {\em IEEE Software}, 29(6):18–21, November 2012.
\newblock URL: \url{http://dx.doi.org/10.1109/ms.2012.167}, \href {https://doi.org/10.1109/ms.2012.167} {\path{doi:10.1109/ms.2012.167}}.

\bibitem{kruchten2012technical}
Philippe Kruchten, Robert~L Nord, and Ipek Ozkaya.
\newblock Technical debt: From metaphor to theory and practice.
\newblock {\em Ieee software}, 29(6):18--21, 2012.

\bibitem{Kruchten2012a}
Philippe Kruchten, Robert~L. Nord, and Ipek Ozkaya.
\newblock Technical debt: {From} metaphor to theory and practice.
\newblock {\em IEEE Software}, 29(6), November 2012.
\newblock URL: \url{http://ieeexplore.ieee.org/document/6336722/}, \href {https://doi.org/10.1109/MS.2012.167} {\path{doi:10.1109/MS.2012.167}}.

\bibitem{li2015systematic}
Zengyang Li, Paris Avgeriou, and Peng Liang.
\newblock A systematic mapping study on technical debt and its management.
\newblock {\em Journal of Systems and Software}, 101:193--220, 2015.

\bibitem{Li2015}
Zengyang Li, Paris Avgeriou, and Peng Liang.
\newblock A systematic mapping study on technical debt and its management.
\newblock {\em Journal of Systems and Software}, 101:193--220, March 2015.
\newblock \href {https://doi.org/10.1016/j.jss.2014.12.027} {\path{doi:10.1016/j.jss.2014.12.027}}.

\bibitem{Martini2018a}
Antonio Martini.
\newblock Anacondebt: {A} tool to assess and track technical debt.
\newblock In {\em Proceedings - {International} {Conference} on {Software} {Engineering}}, 2018.
\newblock \href {https://doi.org/10.1145/3194164.3194185} {\path{doi:10.1145/3194164.3194185}}.

\bibitem{McConnell2008a}
Steve McConnell.
\newblock {Managing technical debt}.
\newblock {\em Construx Inc.}, 2008.

\bibitem{Mendes2019}
Thiago~S. Mendes, Felipe~G.S. Gomes, David~P. Gonçalves, Manoel~G. Mendonça, Renato~L. Novais, and Rodrigo~O. Spínola.
\newblock {VisminerTD}: a tool for automatic identification and interactive monitoring of the evolution of technical debt items.
\newblock {\em Journal of the Brazilian Computer Society}, 25(1), 2019.
\newblock \href {https://doi.org/10.1186/s13173-018-0083-1} {\path{doi:10.1186/s13173-018-0083-1}}.

\bibitem{nayebi2019longitudinal}
Maleknaz Nayebi, Yuanfang Cai, Rick Kazman, Guenther Ruhe, Qiong Feng, Chris Carlson, and Francis Chew.
\newblock A longitudinal study of identifying and paying down architecture debt.
\newblock In {\em 2019 IEEE/ACM 41st International Conference on Software Engineering: Software Engineering in Practice (ICSE-SEIP)}, pages 171--180. IEEE, 2019.

\bibitem{Ozkaya2023RTC}
Ipek Ozkaya, Forrest Shull, Julie Cohen, and Brigid O'Hearn.
\newblock Report to the congressional defense committees on natinoal defense authorization act (ndaa) for fiscal year 2022 section 835 independent study on technical debt in software-intensive systems.
\newblock September 2023.
\newblock URL: \url{https://insights.sei.cmu.edu/library/congressional-report-section-835-technical-debt-cmu-sei-2023-tr-003/}.

\bibitem{perera2024systematic}
Judith Perera, Ewan Tempero, Yu-Cheng Tu, and Kelly Blincoe.
\newblock A systematic mapping study exploring quantification approaches to code, design, and architecture technical debt.
\newblock {\em ACM Transactions on Software Engineering and Methodology}, 33(7):1--44, 2024.

\bibitem{pichler2010agile}
Roman Pichler.
\newblock {\em Agile product management with scrum: Creating products that customers love}.
\newblock Pearson Education India, 2010.

\bibitem{Rios2018}
Nicolli Rios, Rodrigo~Oliveira Spínola, Manoel~G. De~Mendonça~Neto, and Carolyn Seaman.
\newblock A study of factors that lead development teams to incur technical debt in software projects.
\newblock In {\em Proceedings - 44th {Euromicro} {Conference} on {Software} {Engineering} and {Advanced} {Applications}, {SEAA} 2018}, 2018.
\newblock URL: \url{https://ieeexplore.ieee.org/document/8498243/}, \href {https://doi.org/10.1109/SEAA.2018.00076} {\path{doi:10.1109/SEAA.2018.00076}}.

\bibitem{Rozanski2011}
Nick Rozanski and E\'{o}in Woods.
\newblock {\em Software Systems Architecture: Working With Stakeholders Using Viewpoints and Perspectives, 2nd Edition}.
\newblock Addison-Wesley Professional, 2011.

\bibitem{sarkar2006api}
Santonu Sarkar, Girish~Maskeri Rama, and Avinash~C Kak.
\newblock Api-based and information-theoretic metrics for measuring the quality of software modularization.
\newblock {\em IEEE Transactions on Software Engineering}, 33(1):14--32, 2006.

\bibitem{sas2022evolution}
Darius Sas, Paris Avgeriou, and Umut Uyumaz.
\newblock On the evolution and impact of architectural smells—an industrial case study.
\newblock {\em Empirical Software Engineering}, 27(4):86, 2022.

\bibitem{schmid_formal_2013}
Klaus Schmid.
\newblock A formal approach to technical debt decision making.
\newblock In {\em Proceedings of the 9th international {ACM} {Sigsoft} conference on {Quality} of software architectures {(QoSA'13)}}, pages 153--162, New York, NY, USA, June 2013. Association for Computing Machinery.
\newblock URL: \url{https://dl.acm.org/doi/10.1145/2465478.2465492}, \href {https://doi.org/10.1145/2465478.2465492} {\path{doi:10.1145/2465478.2465492}}.

\bibitem{seaman2011measuring}
Carolyn Seaman and Yuepu Guo.
\newblock Measuring and monitoring technical debt.
\newblock In {\em Advances in Computers}, volume~82, pages 25--46. Elsevier, 2011.

\bibitem{Sharma2018}
Tushar Sharma and Diomidis Spinellis.
\newblock A survey on software smells.
\newblock {\em Journal of Systems and Software}, 138:158 -- 173, 2018.
\newblock \href {https://doi.org/10.1016/j.jss.2017.12.034} {\path{doi:10.1016/j.jss.2017.12.034}}.

\bibitem{sierra2019survey}
Giancarlo Sierra, Emad Shihab, and Yasutaka Kamei.
\newblock A survey of self-admitted technical debt.
\newblock {\em Journal of Systems and Software}, 152:70--82, 2019.

\bibitem{SonarQube}
{SonarQube}.
\newblock URL: \url{https://www.sonarqube.org/downloads/}.

\bibitem{Tan2023}
Jie Tan, Daniel Feitosa, and Paris Avgeriou.
\newblock The lifecycle of technical debt that manifests in both source code and issue trackers.
\newblock {\em Information and Software Technology}, 159:107216, July 2023.
\newblock URL: \url{http://dx.doi.org/10.1016/j.infsof.2023.107216}, \href {https://doi.org/10.1016/j.infsof.2023.107216} {\path{doi:10.1016/j.infsof.2023.107216}}.

\bibitem{TeamScale}
{TeamScale}.
\newblock URL: \url{https://teamscale.com/}.

\bibitem{Tonin2017}
Graziela~Simone Tonin, Alfredo Goldman, Carolyn Seaman, and Diogo Pina.
\newblock Effects of {Technical} {Debt} {Awareness}: {A} {Classroom} {Study}.
\newblock In Hubert Baumeister, Horst Lichter, and Matthias Riebisch, editors, {\em 18th {International} {Conference}, {XP} 2017}, 2017.
\newblock URL: \url{http://link.springer.com/10.1007/978-3-319-57633-6}, \href {https://doi.org/10.1007/978-3-319-57633-6_6} {\path{doi:10.1007/978-3-319-57633-6_6}}.

\bibitem{Verdecchia2020}
Robert Verdeccia, Philippe Kruchten, Patricia Lago, Roberto Verdecchia, Philippe Kruchten, and Patricia Lago.
\newblock Architectural {Technical} {Debt} : {A} {Grounded} {Theory}.
\newblock In A.~Jansen, I.~Malavolta, H.~Muccini, I.~Ozkaya, and O.~Zimmermann, editors, {\em European {Conference} on {Software} {Architecture} ({ECSA} 2020)}. Springer, Cham., 2020.
\newblock \href {https://doi.org/10.1007/978-3-030-58923-3_14} {\path{doi:10.1007/978-3-030-58923-3_14}}.

\bibitem{wiese_it_2023}
Marion Wiese and Klara Borowa.
\newblock {IT} managers’ perspective on {Technical} {Debt} {Management}.
\newblock {\em Journal of Systems and Software}, 202:111700, April 2023.
\newblock \href {https://doi.org/10.1016/j.jss.2023.111700} {\path{doi:10.1016/j.jss.2023.111700}}.

\bibitem{Wiese2022}
Marion Wiese, Paula Rachow, Matthias Riebisch, and Julian Schwarze.
\newblock Preventing technical debt with the {TAP} framework for {Technical} {Debt} {Aware} {Management}.
\newblock {\em Information and Software Technology}, 2022.
\newblock URL: \url{https://www.sciencedirect.com/science/article/pii/S0950584922000787}, \href {https://doi.org/10.1016/j.infsof.2022.106926} {\path{doi:10.1016/j.infsof.2022.106926}}.

\bibitem{yli-huumo_sources_2014}
Jesse Yli-Huumo, Andrey Maglyas, and Kari Smolander.
\newblock The sources and approaches to management of technical debt: {A} case study of two product lines in a middle-size finnish software company.
\newblock In {\em 15th {International} {Conference} on {Product}-{Focused} {Software} {Process} {Improvement}, {PROFES} 2014}, volume 8892, 2014.
\newblock ISSN: 16113349.
\newblock \href {https://doi.org/10.1007/978-3-319-13835-0_7} {\path{doi:10.1007/978-3-319-13835-0_7}}.

\bibitem{Yli-Huumo2016}
Jesse Yli-Huumo, Andrey Maglyas, and Kari Smolander.
\newblock How do software development teams manage technical debt? – {An} empirical study.
\newblock {\em Journal of Systems and Software}, 120, 2016.
\newblock \href {https://doi.org/10.1016/j.jss.2016.05.018} {\path{doi:10.1016/j.jss.2016.05.018}}.

\end{thebibliography}
